\documentclass[conference]{IEEEtran}
\IEEEoverridecommandlockouts
\usepackage{cite}
\usepackage{amsmath,amssymb,amsfonts}
\usepackage{algorithmic}
\usepackage{graphicx}
\usepackage{textcomp}
\usepackage{xcolor}
\usepackage{url}
\def\BibTeX{{\rm B\kern-.05em{\sc i\kern-.025em b}\kern-.08em
    T\kern-.1667em\lower.7ex\hbox{E}\kern-.125emX}}
\begin{document}

\title{Uplink SRS-Based Real-Time Indoor Localization System over OpenAirInterface}

\author{\IEEEauthorblockN{
        Ping-Yu Hsieh\IEEEauthorrefmark{1},
        Chieh-Chun Chen\IEEEauthorrefmark{2},
        Navid Nikaein\IEEEauthorrefmark{2}\IEEEauthorrefmark{3}, and
        Ray-Guang Cheng\IEEEauthorrefmark{1}
        \\
  }
 \IEEEauthorblockA{\IEEEauthorrefmark{1}
Dept. of Electronic and Computer Engineering, National Taiwan Univ. of Science and Technology, Taiwan
  }
 \IEEEauthorblockA{\IEEEauthorrefmark{2}
 BubbleRAN, France}
\IEEEauthorblockA{\IEEEauthorrefmark{3}
 Communication Systems Department, EURECOM, France}

Email: navid.nikaein@eurecom.fr, crg@mail.ntust.edu.tw
}

\maketitle
\begin{abstract}
Indoor localization is one of the important services for future 5G-Advanced and 6G systems. This paper presents an uplink Sounding Reference Signal (SRS)-based real-time indoor localization system implemented over an OpenAirInterface (OAI) 5G Radio Access Network (RAN).
The proposed system uses a Positioning xApp to derive Channel Frequency Response (CFR) measurements from uplink SRS measurements. The SRS measurements are obtained from the gNB through the E2 Service Model for Lower Layer Control (E2SM-LLC) over the standardized E2 interface. The xApp transforms the CFR into a 32-dimensional physics-aware feature vector and uses a Random Forest (RF) regressor to estimate the two-dimensional position of the user equipment.
We implemented the Positioning xApp on an OAI-based 5G testbed in a multipath-rich indoor laboratory at EURECOM to validate the proposed system. Experimental results show that the proposed system achieves a mean absolute error (MAE) of 0.12 m under random train-test evaluation. These results demonstrate the feasibility and limitations of uplink SRS-based real-time indoor localization over OAI.

\end{abstract}


\section{Introduction}

Indoor localization is an important service for future 5G-Advanced and 6G networks. Recent studies have shown that communication signals can also be used for sensing and positioning applications. At the same time, the O-RAN architecture enables developers to deploy intelligent applications as xApps on the Near-RT RIC platform.

Several localization methods have been proposed. RSSI-based approaches are simple to implement but often suffer from poor accuracy in indoor environments due to multipath propagation. More advanced solutions use deep learning models and channel measurements to improve positioning performance. However, these methods usually require higher computational complexity.

In this paper, we implement a real-time indoor localization system on an OpenAirInterface (OAI) based O-RAN testbed. The proposed system collects uplink Sounding Reference Signal (SRS) measurements from the gNB through the E2 interface and converts them into Channel Frequency Response (CFR) data. A Positioning xApp extracts a compact feature vector from the CFR and uses a Random Forest (RF) regressor to estimate the user position.

The main contributions of this work are summarized as follows:

\begin{itemize}
\item Implementation of an end-to-end uplink SRS-based localization system on an OAI O-RAN platform.
\item Design of a lightweight feature extraction method using delay, phase, and received power information.
\item Experimental validation in a real indoor environment using commercial user equipment.
\item Demonstration of real-time single-UE and multi-UE localization visualization.
\end{itemize}

The rest of the paper is organized as follows. Section \ref{SysMod} introduces the system model. Section \ref{Positioning} details the proposed Positioning xApp and the feature extraction pipeline. Section \ref{result} presents the experimental results. The conclusion is given in Section \ref{conclusion}.

\section{System Model}\label{SysMod}
\begin{figure}[htbp]
    \centering
    \includegraphics[width=0.6\linewidth]{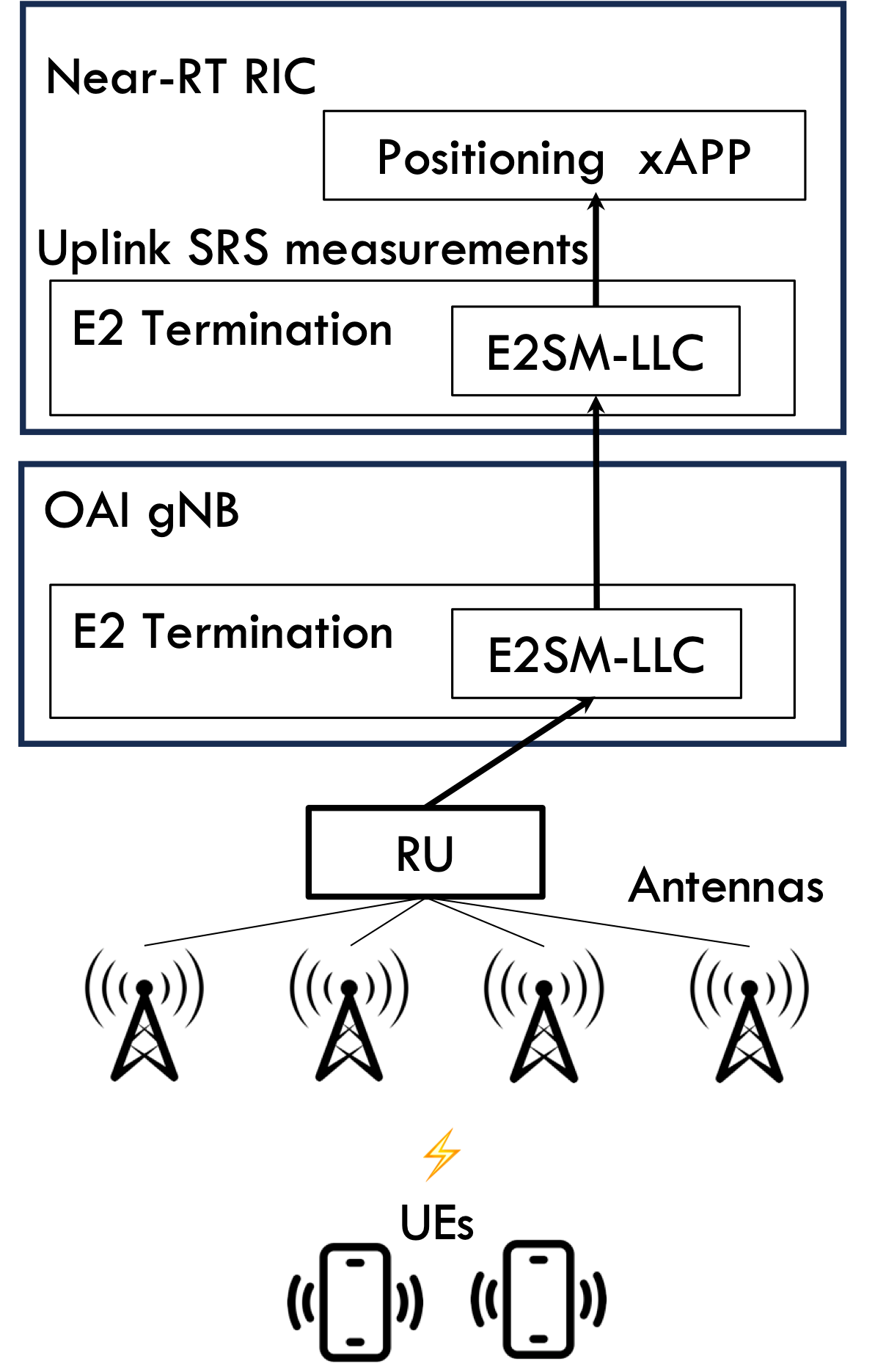}
    \vspace{-2mm}
    \caption{System architecture of the proposed uplink SRS-based real-time indoor localization system.}
    \label{fig:env}
    \vspace{-3mm}
\end{figure}

The proposed ISAC system is built on an end-to-end 5G-compliant O-RAN testbed, as illustrated in Fig. \ref{fig:env}. The foundational infrastructure consists of a 5G Core (Open5GS), a software-defined gNodeB (OpenAirInterface) \cite{oai_5g}, and a 4-antenna Radio Unit (RU). Commercial User Equipment (UE), such as a standard smartphone (e.g., Pixel 7), acts as the target device, continuously transmitting uplink Sounding Reference Signals (SRS) over the air during normal communication operations.

To enable passive data extraction without disrupting active communications, an SRS Data Collecting xApp is deployed on the Near-RT RIC. Through the E2 interface (managed by the E2 Termination function) and the Low Layer Control Service Model (LLC SM), the gNB streams real-time SRS measurements to the RIC. This raw SRS telemetry is then forwarded to the Positioning xApp for spatial intelligence processing.

The objective of the system is to estimate the 2D position of a UE from uplink SRS measurements. Let the CFR matrix extracted from the 4-antenna grid be defined as $\mathbf{H} \in \mathbb{C}^{4 \times N}$, where $N$ is the number of SRS subcarriers allocated. Let $H_i[k]$ represent the complex I/Q sample for the $i$-th antenna on the $k$-th subcarrier.

The system is designed to drive a real-time ISAC spatial positioning application. The operational objective is to derive the 2D physical coordinates $\mathbf{y} = (x, y)$ of the UE transmitting within the indoor environment directly from the input $\mathbf{H}$. The localization problem can be expressed as $f: \mathbf{H} \rightarrow \mathbf{y}$ that operates under strict microsecond-level latency constraints, balancing both high spatial precision and lightweight computational overhead.

\section{Positioning xApp}\label{Positioning}
\begin{figure}[htbp]
    \centering
    \includegraphics[trim=0 35 0 0, clip, width=1.2\linewidth]{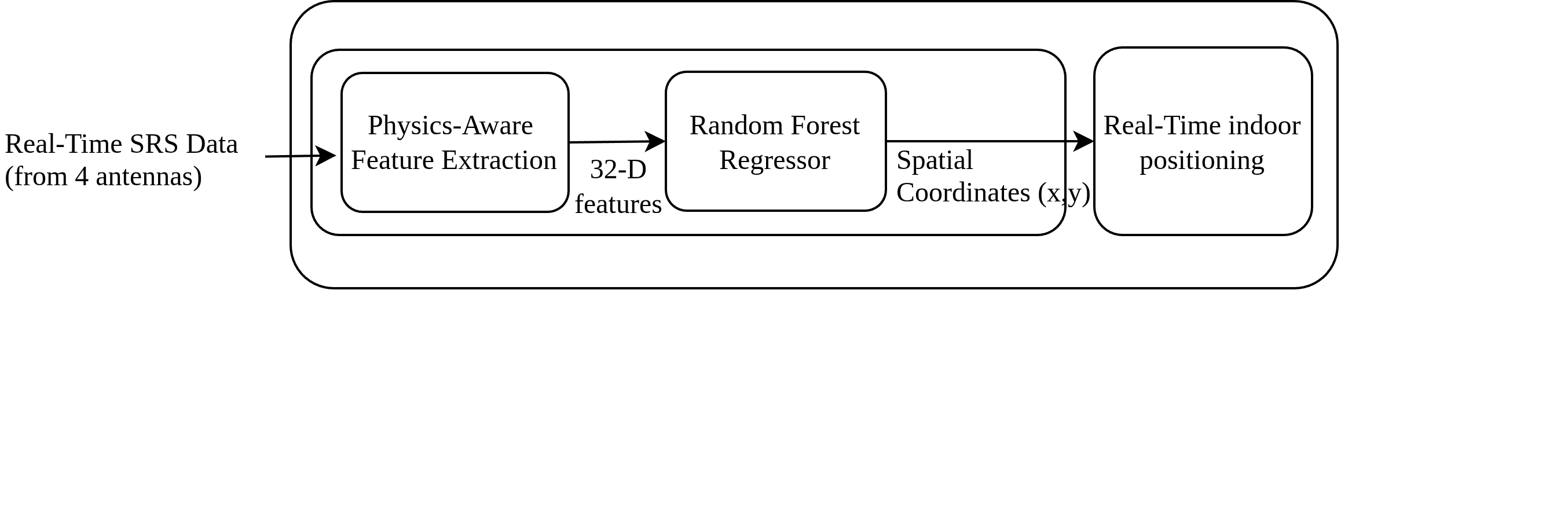}
    \caption{Functional block diagram of the proposed Positioning xApp.}
    \label{fig:method}
\end{figure}

Figure~\ref{fig:method} shows the functional block diagram of the proposed Positioning xApp. The xApp consists of a lightweight, two-stage inference pipeline.

Let the raw CFR for the $i$-th antenna be $H_i[k] = I_i[k] + jQ_i[k]$. Since directly utilizing raw CFR is sub-optimal due to hardware phase noise, we map these signals into a robust 32-D spatial feature vector.

\noindent\textbf{16-D Normalized Power Delay Profile:} To capture multipath propagation, we transform the frequency-domain CFR into the time-domain Channel Impulse Response (CIR) via an $N$-point IFFT:
\begin{equation}
h_i[n] = \frac{1}{N} \sum_{k=0}^{N-1} H_i[k] e^{j2\pi kn/N}
\end{equation}
We compute the Power Delay Profile (PDP) and align it to the peak reference tap $n_{ref}$. The spatial-averaged PDP across 4 antennas is normalized by its global maximum:
\begin{equation}
\tilde{P}[n] = \frac{\frac{1}{4} \sum_{i=1}^{4} |h_i[n - n_{ref}]|^2}{\max \left( \frac{1}{4} \sum_{i=1}^{4} |h_i[n - n_{ref}]|^2 \right)}
\end{equation}
We truncate this profile to the first 16 time-domain taps, which is sufficient to capture the dominant multipath energy within the indoor environment.

\noindent\textbf{8-D Inter-Antenna Phase Difference:} To approximate Angle of Arrival (AoA) signatures, we compute the phase difference $\Delta\phi_i$ between the $i$-th receiving antenna and the reference antenna at the peak tap $n_{ref}$. To prevent phase wrapping discontinuities, features are encoded trigonometrically as $\mathbf{f}_{\text{phase}, i} = [\cos(\Delta\phi_i), \sin(\Delta\phi_i)]$, yielding 8 orientation features.

\noindent\textbf{8-D Received Power and Spatial Symmetry:} Distance-dependent path loss is represented by the received energy (dB) with a hardware compensation offset $C_i$:
\begin{equation}
P_{i, \text{dB}} = 10 \log_{10} \left( \frac{1}{N} \sum_{k=0}^{N-1} |H_i[k]|^2 + \epsilon \right) + C_i
\end{equation}
where $\epsilon = 10^{-9}$. Finally, we extract 4 spatial heuristics based on the power profiles: global mean RSSI, maximum RSSI, and the differential power between opposite antenna pairs ($\Delta X, \Delta Y$) to capture intrinsic geometric orientation.

We then employ a Random Forest (RF) regressor to map the high-dimensional, non-linear radio fingerprints to 2D physical coordinates $\mathbf{y} = (x, y)$.  To optimize localization stability and inference speed, the regressor is structured with $K=100$ independent decision trees, a maximum depth constraint of $max\_depth=12$ to prevent overfitting to environmental fluctuations, and a minimum sample split threshold of $min\_samples\_split=20$ to guarantee spatial generalization. These parameters were selected to balance localization accuracy and computational complexity. 

During real-time inference, the final estimated coordinate is derived by averaging the continuous spatial predictions from all $K$ trees:
\begin{equation}
\hat{\mathbf{y}} = \frac{1}{K} \sum_{k=1}^{K} T_k(\mathbf{x})
\end{equation}
This ensemble mechanism intrinsically filters out extreme spatial outliers caused by transient blockages. The prediction variance serves as an explicit indicator of localization uncertainty, driving the Probability Density Heatmap.

The final stage of the proposed pipeline translates the raw spatial coordinates $\hat{\mathbf{y}}$ into actionable visual intelligence via a real-time dashboard. Rather than rendering a simple deterministic point, the application leverages the prediction variance from the RF ensemble to project a dynamic probability density heatmap. This approach visually represents the localization confidence, naturally highlighting areas of high multipath uncertainty.

We apply an Exponential Moving Average (EMA) filter to the output coordinates. This filter reduces trajectory fluctuations and screen jitter. Although the filter adds a slight visual delay, it stabilizes the tracking trajectory. Thus, the dashboard accurately reflects human walking patterns. This completes the end-to-end ISAC pipeline, bridging raw O-RAN telemetry to an intuitive spatial awareness interface.

\section{Experimental Results}\label{result}

We evaluated the system in a $6 \times 6$ m indoor laboratory. Glass partitions and concrete walls in this room cause severe multipath reflections. We collected an extensive dataset of over 23,000 SRS frames across dense reference grid points. For the spatial inference model, the Random Forest regressor is configured with $K=100$ estimators, $max\_depth=12$, and $min\_samples\_split=20$, striking a balance between spatial resolution and microsecond-level inference latency. 

Under a standard 80/20 random split, the proposed model achieves a Mean Absolute Error (MAE) of 0.1238~m. To assess the robustness of the proposed method beyond interpolation within the training distribution, we conduct a spatially separated blind evaluation. In this setting, specific coordinate regions are completely excluded from the training process and used exclusively for testing. The results show that the MAE increases to 1.78~m, indicating a significant degradation when the model is required to extrapolate to unseen spatial regions. This behavior highlights the inherent limitation of tree-based regressors, which rely on discrete decision partitions and lack continuous spatial interpolation capability.

\begin{figure}[htbp]
    \centering
    \includegraphics[width=1\linewidth]{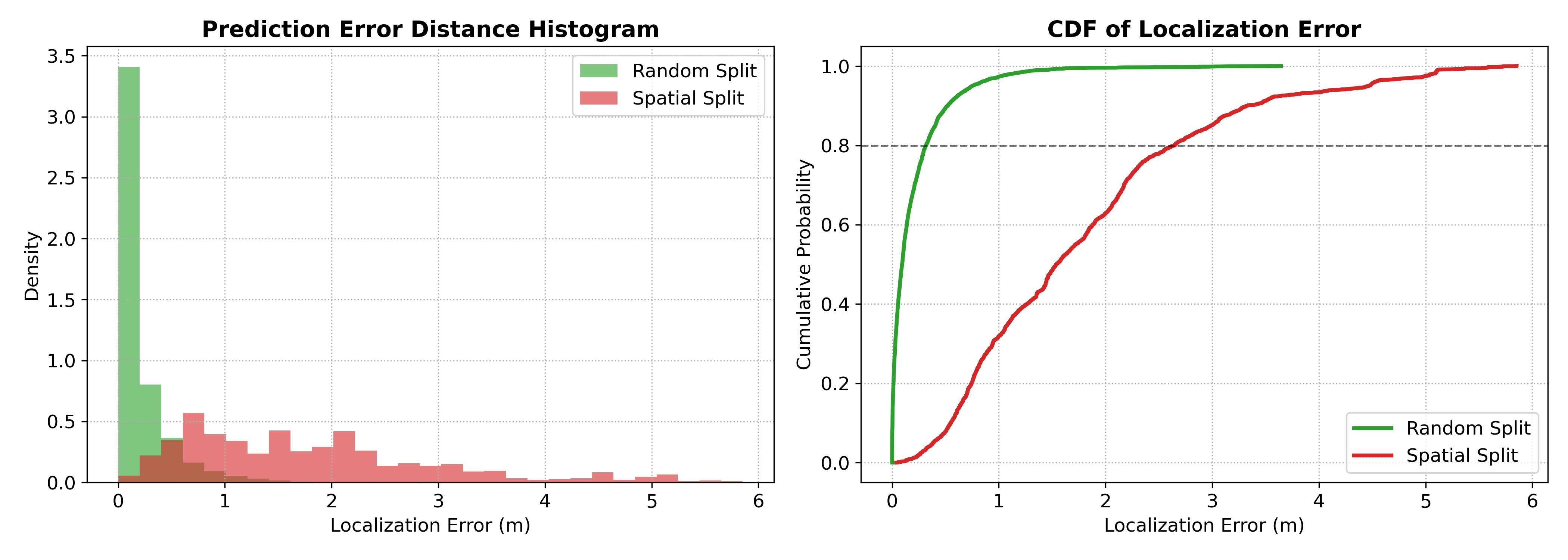}
    \vspace{-2mm}
    \caption{Prediction error histogram and CDF under standard and spatial split evaluation.}
    \label{fig:error_hist}
    \vspace{-3mm}
\end{figure}

To validate the effectiveness of the proposed physics-aware representation, we further evaluate a lightweight baseline using only the Received Signal Strength Indicator (RSSI) from the four antennas, which produces a significantly degraded MAE of 0.2153~m. This corresponds to a 42\% reduction in localization error, demonstrating that the incorporation of the delay and phase-domain structure is essential for accurate indoor positioning under multipath conditions.
As depicted in the CDF of the localization error (Fig. \ref{fig:error_hist}), the system achieves approximately 85\% of predictions within a 0.5 m error threshold.

We also developed a dashboard to visualize the instantaneous location along a 3-second historical trajectory. Rather than rendering a deterministic point estimate, the system projects a dynamic probability heatmap to reflect localization confidence. To mitigate trajectory fluctuations induced by unstable RF signals, an Exponential Moving Average (EMA) filter is applied. Although this temporal smoothing introduces a marginal processing lag of approximately 1 second, it effectively stabilizes the visual output and remains well within the acceptable tolerance for tracking human walking speeds.

\begin{figure}[htbp]
    \centering
    \includegraphics[width=1\linewidth]{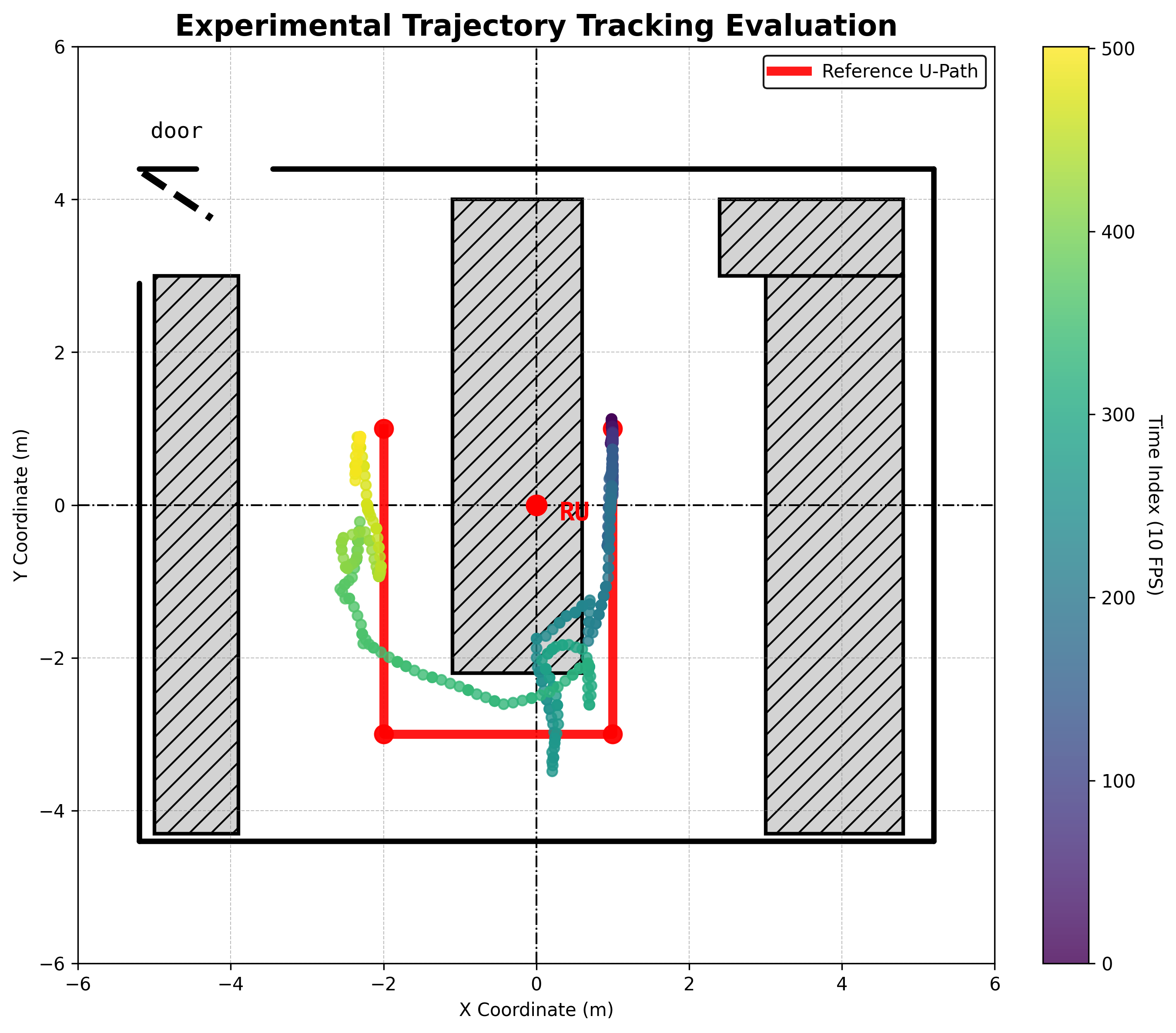}
    \vspace{-2mm}
    \caption{Experimental real-time trajectory tracking. Central open zones exhibit excellent alignment, while the bottom glass partitions introduce multipath scattering.}  
    \label{fig:trajectory}
    \vspace{-3mm}
\end{figure}

As demonstrated in Fig. \ref{fig:trajectory}, trajectory tracking captures UE movement over a 50-second interval. Spatial correlation indicates that tracking deviations at the bottom boundary are primarily induced by close proximity to the room's glass partitions, which act as reflectors generating intense multipath scattering.

\begin{figure}[htbp]
    \centering
    \includegraphics[width=1\linewidth]{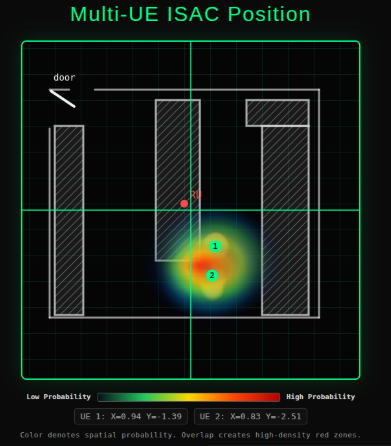}
    \vspace{-2mm}
    \caption{Real-time Multi-UE spatial tracking dashboard. Overlapping probability halos automatically highlight high-density collision zones.}
    \label{fig:multiue}
    \vspace{-3mm}
\end{figure}

When tracking multiple UEs, rather than outputting deterministic coordinates, the dashboard projects individual continuous location probability maps onto the shared grid. As Fig. \ref{fig:multiue} shows, when multiple UEs physically converge, their spatial probability distributions geometrically superimpose into a high-density red core, visually detecting collisions without additional computational overhead.

\section{Conclusion}\label{conclusion}

In this paper, we presented a real-time indoor localization system deployed on a commercial O-RAN testbed. By leveraging robust feature extraction and a Random Forest regressor, the system successfully uses uplink SRS measurements for indoor localization, enabling sub-meter active positioning (0.12~m MAE) while remaining computationally lightweight.

While the system is evaluated in a single indoor environment, this design is intentional, as the proposed lightweight feature extraction and training pipeline are designed for rapid re-deployment in new environments. This enables efficient data collection and model retraining with minimal overhead when transitioning to different deployment sites.

Despite these achievements, specific challenges remain. Complex multipath reflections near room boundaries occasionally induce localization instability and prediction jumps. The EMA filter tracks pedestrian movement with low latency. However, smoothing the signal jitter slightly increases the system response time.

To address these limitations, future work will explore deep learning architectures such as CNN-based and transformer-based models to better capture spatial correlations in multipath environments. In addition, adaptive filtering strategies will be investigated to improve robustness under varying mobility conditions.
A live video demonstration of our real-time ISAC dashboard is available online \cite{isac_demo}.

\section*{ACKNOWLEDGEMENT}
This work was supported by the National Science and Technology Council (NSTC), Taiwan, under Contract numbers 112-2218-E-011-006 and 114-2221-E-011 -078 -MY3.


\begin{thebibliography}{00}

\bibitem{isac_survey}
F. Liu et al., ``Integrated Sensing and Communications: Toward Dual-Functional Wireless Networks for 6G and Beyond,'' \emph{IEEE JSAC}, vol. 40, no. 6, pp. 1728-1767, 2022.

\bibitem{oran_xapp}
M. Polese et al., ``Understanding O-RAN: Architecture, Interfaces, Algorithms, Security, and Research Challenges,'' \emph{IEEE COMST}, vol. 25, no. 2, pp. 1376-1411, 2023.

\bibitem{otani_sdr}
H. Otani et al., ``An Open-Source SDR-Based Device-Free Sensing Platform for ISAC,'' \emph{IEEE OJ-COMS}, vol. 6, pp. 9982-9990, 2025.

\bibitem{carbonara_downlink}
S. Carbonara et al., ``Downlink ISAC with a Full-Stack 5G Experimental Testbed,'' in \emph{IEEE JC\&S}, 2026, pp. 1-6.

\bibitem{bouknana_oran}
N. Bouknana et al., ``An O-RAN Framework for AI/ML-Based Localization with OpenAirInterface and FlexRIC,'' in \emph{Proc. WONS}, 2026.

\bibitem{rssi_radar}
P. Bahl and V. N. Padmanabhan, ``RADAR: An in-building RF-based user location and tracking system,'' in \emph{Proc. IEEE INFOCOM}, vol. 2, 2000, pp. 775-784.

\bibitem{3gpp_pos}
3GPP, ``Study on NR positioning support,'' TS 38.855 v16.0.0, 2019.

\bibitem{mundlamuri_tools}
R. Mundlamuri et al., ``5G NR Positioning with OpenAirInterface: Tools and Methodologies,'' in \emph{Proc. WONS}, 2025.

\bibitem{oai_5g}
F. Kaltenberger et al., ``OpenAirInterface: Democratizing innovation in the 5G era,'' \emph{Computer Networks}, vol. 176, p. 107284, 2020.

\bibitem{isac_demo}
BubbleRAN, ``5G ISAC Spatial Intelligence,'' YouTube, 2026. [Online]. Available: \url{https://youtu.be/hAeIbw2aTQQ}

\end{thebibliography}
\end{document}